\documentclass[10pt,conference]{IEEEtran}
\IEEEoverridecommandlockouts
\usepackage{cite}
\usepackage{amsmath,amssymb,amsfonts}
\usepackage{algorithmic}
\usepackage{graphicx}
\usepackage{textcomp}
\usepackage{xcolor}
\def\BibTeX{{\rm B\kern-.05em{\sc i\kern-.025em b}\kern-.08em
    T\kern-.1667em\lower.7ex\hbox{E}\kern-.125emX}}

\usepackage{amssymb}
\usepackage{hyperref}
\usepackage[plain]{fancyref}
\usepackage{ifdraft}

\usepackage[inline]{enumitem}
\usepackage{xcolor}
\usepackage{xspace}
\usepackage[final]{listings}
\usepackage{acronym}
\usepackage{url}
\usepackage{amsmath}
\usepackage{amssymb}
\usepackage{booktabs} 
\usepackage{subfig}
\usepackage{balance}
\usepackage{dirtree}

\usepackage[ruled]{algorithm2e} 

\newcommand{\subparagraph}{}
\usepackage[compact]{titlesec}
\titlespacing{\section}{0pt}{*0}{*0}
\titlespacing{\subsection}{0pt}{*0}{*0}
\titlespacing{\subsubsection}{0pt}{*0}{*0}

\usepackage{etoolbox}
\makeatletter
\patchcmd{\@makecaption}
  {\scshape}
  {}
  {}
  {}
\patchcmd{\@makecaption}
  {\\}
  {.\ }
  {}
  {}
\makeatother

\definecolor{OliveGreen}{rgb}{0,0.6,0.3}

\renewcommand{\lstlistingname}{Snippet}
\newcommand*{\fancyreflstlabelprefix}{lst}
\newcommand*{\Freflstname}{\lstlistingname}
\newcommand*{\freflstname}{\MakeLowercase{\lstlistingname}}
\Frefformat{vario}{\fancyreflstlabelprefix}%
  {\Freflstname\fancyrefdefaultspacing#1#3}
\frefformat{vario}{\fancyreflstlabelprefix}%
  {\freflstname\fancyrefdefaultspacing#1#3}
\Frefformat{plain}{\fancyreflstlabelprefix}%
  {\Freflstname\fancyrefdefaultspacing#1}
\frefformat{plain}{\fancyreflstlabelprefix}%
  {\freflstname\fancyrefdefaultspacing#1}

\renewcommand{\tablename}{Table}  
  
\newcommand*{\fancyreflnlabelprefix}{ln}
\newcommand*{\Freflnname}{Line}
\newcommand*{\freflnname}{\MakeLowercase{\Freflnname}}
\Frefformat{vario}{\fancyreflnlabelprefix}%
  {\Freflnname\fancyrefdefaultspacing#1#3}
\frefformat{vario}{\fancyreflnlabelprefix}%
  {\freflnname\fancyrefdefaultspacing#1#3}
\Frefformat{plain}{\fancyreflnlabelprefix}%
  {\Freflnname\fancyrefdefaultspacing#1}
\frefformat{plain}{\fancyreflnlabelprefix}%
  {\freflnname\fancyrefdefaultspacing#1}

\lstdefinelanguage{JavaScript}{
keywords={typeof, new, true, false, catch, function, return, null, catch, switch, var, if, in, for, while, do, else, case, break, throw, this, instanceof},
keywordstyle=\color{purple}\bfseries,
ndkeywords={},
ndkeywordstyle=\color{blue}\bfseries,
identifierstyle=\color{black},
sensitive=false,
comment=[l]{//},
morecomment=[s]{/*}{*/},
commentstyle=\color{OliveGreen}\ttfamily,
stringstyle=\color{OliveGreen}\ttfamily,
morestring=[b]',
morestring=[b]"
}
\usepackage{color}
\definecolor{gray97}{gray}{.97}
\definecolor{gray90}{gray}{.90}
\definecolor{gray75}{gray}{.75}
\definecolor{gray45}{gray}{.45}
\definecolor{codegreen}{rgb}{0,0.6,0}
\definecolor{codered}{rgb}{0.6,0,0}
\definecolor{codegray}{rgb}{0.5,0.5,0.5}
\definecolor{codepurple}{rgb}{0.58,0,0.82}
\lstdefinestyle{floating}{%
  frame=none,
  float=htb,
  captionpos=b
}

\lstdefinestyle{ctxtraits}
 {language=JavaScript,
  frame=lines,
  showstringspaces=false,
  keywordstyle=\tt\bf,
  tabsize=3,
  style=floating,
  morekeywords={Trait, cop, Context, activate, deactivate, adapt, addObjectPolicy, manager}
}

\lstnewenvironment{ctxtraits}[1][]
 {\lstset{style=ctxtraits,#1}}{}

\newcommand{\eg}{\emph{e.g.,}\xspace}
\newcommand{\ie}{\emph{i.e.,}\xspace}
\newcommand{\etal}{\emph{et al.}\xspace}

\usepackage{xurl}
\hypersetup{breaklinks=true}
\usepackage[hyphenbreaks]{breakurl}

\definecolor{author}{rgb}{.5, .5, .5}
\definecolor{comment}{rgb}{.1, .0, .9}
\definecolor{note}{rgb}{.9, .4, .0}
\definecolor{idea}{rgb}{.1, .7, .0}
\definecolor{missing}{rgb}{.9, .1, .0}
\definecolor{deleteme}{rgb}{.9, .1, .0}

\acrodef{APK}{Android Application Package}

\begin{document}

\title{Towards machine learning guided by best practices}

\author{\IEEEauthorblockN{Anamaria Mojica-Hanke}
\IEEEauthorblockA{\textit{University of Passau} \\
Passau, Germany}
\IEEEauthorblockA{\textit{Universidad de los Andes} \\
	Bogota, Colombia\\
	ai.mojica10@uniandes.edu.co}

}

\maketitle

\begin{abstract}

Nowadays, machine learning (ML) is being used in software systems with multiple application fields, from medicine to software engineering (SE). On the one hand, the popularity of ML in the industry can be seen in the statistics showing its growth and adoption.  On the other hand, its popularity can also be seen in research, particularly  in SE, where multiple studies related to the use of Machine Learning in Software Engineering have been published in conferences and journals.  At the same time, researchers and practitioners have shown that machine learning has some particular challenges and pitfalls. In particular, research has shown that ML-enabled systems  have a different development process than traditional software, which also describes  some of the challenges of ML applications.  In order to mitigate some of the identified challenges and pitfalls, white and gray literature has proposed a set of recommendations based on their own experiences  and focused on their domain (\eg biomechanics), but for the best of our knowledge, there is no guideline focused on the SE community. 
This thesis aims to reduce the gap of not having clear guidelines in the SE community by using  possible sources of practices such as question-and-answer communities and also previous research studies. As a result, we will  present a set of practices with an SE perspective, for researchers and practitioners, including a tool for searching them. 

\end{abstract}


\section{Motivation - Problem definition}
\label{sec:motiv}

Machine learning (ML) has multiple fields of applications,  such as finance, medicine, education,  and software engineering. Indeed, in each  of these fields, ML has different kinds of applications; for example, fraud detection~\cite{awoyemi2017credit} and trading~\cite{sebastiao2021forecasting};  cancer detection~\cite{saba2020recent} and outbreak prediction~\cite{ardabili2020covid};  program translation~\cite{roziere2020unsupervised}; and code transformation~\cite{tufano2019learning}. Showing the wide adoption of ML for different tasks and disciplines and its capability to affect  multiple domains. 

The aforementioned  impact could also be seen in the industry, as it has not only grown in its usage demand but also its popularity and benefits.  For instance, a \textit{McKinsey} report by Chui \etal states that 50\% of the respondents to their study answered that their companies had adopted artificial intelligence  in at least one business function~\cite{chui_hall_mayhew_singla_2022};  this is also supported by the most recent \textit{NewVantage Partner 2022 Data and AI Executive survey (NPAIS)}, that  shows that 91\% of organizations are investing in AI activities~\cite{bean_2022}. Complementary,  on the same survey, NPAIS, it is reported that 92.1\% of organizations are realizing measurable benefits~\cite{bean_2022}, which could also be seen in the survey realized by  \textit{McKinsey} in which it is reported that AI has helped in increasing the revenue of companies in sales and product development~\cite{chui_hall_mayhew_singla_2022}.

On the not-that-bright side,  recent studies have shown that ML-enabled systems (\ie systems that have at least one ML component)  have  challenges~\cite{Alshangiti_2019}, pitfalls~\cite{bone2015applying, biderman2020pitfalls, MichaelLones2021}, problems~\cite{sculley2015hidden}, or mismatches~\cite{LewisGrace2021WAIN} as any software development process and system. Some studies also have shown that  ML challenges and problems are also  publicly discussed in communities  such as Stack Overflow~\cite{Alshangiti_2019, Islam_2019, hamidi2021towards}. Moreover, some studies indicate that ML systems have particular problems and challenges~\cite{LewisGrace2021WAIN, sculley2015hidden}, and could be related to inadequate documentation and communication across the different actors involved in the ML development~\cite{LewisGrace2021WAIN}; or technical debt~\cite{sculley2015hidden}.   In addition,  the development of ML systems involves different phases from the traditional software ones~\cite{amershi2019software}, and in each of those phases, different challenges could be presented~\cite{LewisGrace2021WAIN, Alshangiti_2019, hamidi2021towards}.

Concerning ML challenges from an industry perspective, some surveys show common fears and challenges faced by companies.  Including problems  when collecting data\cite{creg_baker_2019}, data quality~\cite{creg_baker_2019}, versioning, and reproducibility of the models~\cite{thormundsson_2022}.  There are also some risks associated with artificial intelligence, such as equity and fairness or personal/individual privacy, that companies are working to mitigate~\cite{chui_hall_mayhew_singla_2022}.

In order to avoid, mitigate  or deal with the  ML challenges, pitfalls, and problems, some studies have proposed a series of recommended guidelines  and best practices based on their own experience and focused on their respective discipline, \eg~\cite{biderman2020pitfalls, MichaelLones2021, halilaj2018machine}. Additionally, there is a plethora of publications (\eg books, blogs) on the field of ML; for example, grey literature, such as the Google  article by Zinkevich~\cite{zinkevich_2021}, is  publicly available and could be considered as a first step with general practices derived from anecdotal evidence. However,  to the best of our knowledge, there are no handbooks listing best practices for using ML focused on SE practitioners or researchers, \ie software engineers and researchers. As ML is becoming more and more involved in SE development projects, bad practices should be avoided to prevent inadequate model planning, implementation, tuning, testing, deployment, and monitoring of ML implementations, \eg \cite{ LewisGrace2021WAIN, amershi2019software}.  The interest in ML has also been displayed in the SE community as more workshops, and conferences related to ML are being colocated within SE conferences, \eg~\cite{furlinger_2023, cain_2022, icse_2022, A2C2_2022, MaLTeSQuE_2022, EASEAI_2022}.  We aim to  reduce the gap of not having the aforementioned handbook by (i) studying what the best practices discussed by practitioners are; (ii) analyzing what  the practices used by SE researchers when executing studies that involve ML  models are, and (iii) building a taxonomy and a handbook of ML practices that could be used by the SE community that uses ML in order to guide their ML development and research.



\section{Related Work}

As previously mentioned, there is a plethora of ML literature; this literature can be peer-reviewed (\ie white literature) or not (\eg technical reports and blog post)~\cite{soldani2019grey}. 

\subsection{Gray literature}
A quick  search on Google using the keywords  ``machine learning'' yields more than 647 million results; if we further specify  the query to include ``machine learning'' AND ``practices,'' the number of hits is reduced to more than 95 million hits, and it gets lower if we search for ``best practices'' instead, which yields about 40 million results instead.  Showing a significant amount of information that could be related to finding good guidelines when developing an ML system. 

Some of the gray literature that can be found when searching for ML practices or guidelines are published by recognized institutions such as SAS~\cite{wujek2016best}, Google~\cite{goole_pair, zinkevich_2021},  or Carnegie Mellon~\cite{horneman2020ai}. In particular, the aforementioned works present a list of practices and guidelines based on the authors' experience. The last two institutions, \cite{zinkevich_2021, horneman2020ai}, present the practices in a broad and general way.  By broad and general,  we mean that the practices are not specific for use cases but general recommendations, such as ``thinking about if the usage of ML/AI is beneficial and necessary or not''. 

Regarding the more detailed practices,  Google~\cite{goole_pair} presents a set of practices related  to designing with AI. The practices are presented in different ways. For example, they are associated with a series of case studies, organized by chapters that are milestones in product development flow, or they are retrieved by questions that guide the practice search. In addition, SAS~\cite{wujek2016best} presents a series of practices that are associated with different stages of the ML development process, \ie data preparation, training, and deployment.  For each of these stages, a brief description of what each stage encompasses is given, followed by a theory of possible approaches, \eg ways to deploy a model. However, this report is missing some  ML pipeline stages like model requirement (\ie the stage  in which  ``designers decide the functionalities that should be included in an ML system, their usefulness for new or existing products''), model evaluation,  data labeling, and model monitoring.

In addition, some of the gray literature covered are pre-prints or papers that are not peer-reviewed, \eg~\cite{MichaelLones2021, Clemmedsson2018}. Lones \etal~\cite{MichaelLones2021} present a series of challenges that they have encountered during their time in academia (\ie teaching and researching). For some of the pitfalls, they present a possible way to handle or avoid the error.  They mainly focus on research and on properties (\ie robustness, reliability) more than stages of the development of the ML system.  Clemmedsson \etal~\cite{Clemmedsson2018}  focus only on ML pitfalls and challenges in an industrial case study where, after a literature review on possible challenges, they surveyed employees in four companies. As a result, they  identify that some challenges are similar to the traditional SE ones, but some of them are only ML-specific challenges, showing that not all the ML challenges can be addressed in the same way as traditional SE challenges and problems.

\subsection{White literature}

A couple of white literature encompassing some best practices in different fields of knowledge that are not targeted for relating ML and SE exists. For instance, \cite{halilaj2018machine} studies pitfalls and challenges in  biomechanics, followed by some practices to deal with them.  In addition, \cite{teschendorff2019avoiding} also focuses  on pitfalls, but in relation to the use of  ML use in omics data science, and they give guidelines to avoid those pitfalls, without giving details  on how those guidelines are extracted. 

Regarding the white literature that discusses ML and SE, there are a couple of studies~\cite{tantithamthavorn2018experience, breck2017ml, amershi2019software, serban2020adoption,ArpQuiPen_22} but with a different approach than the desired one (\ie ML for SE). Some areas of SE, such as defect modelling, use ML as a tool to achieve its purpose, which in this case would be to \textit{``understand actionable insights in order to make better decisions related to the different phases of software practice''}~\cite{tantithamthavorn2018experience}. And associated with ML use, research has been conducted to find and understand some problems and challenges, also giving active recommendations on how to handle them~\cite{tantithamthavorn2018experience}. However, those guidelines do not only focus on ML but also on different aspects that are associated with the specific ML applications without a clear separation of those two aspects. 
In addition,  those guides are not always associated with other possible SE applications in which they can be applied. This  association is also missing in the study by Arp~\etal~\cite{ArpQuiPen_22}, which identifies ten pitfalls in learning-based security systems and  evaluates their existence in the security literature using ML. They also give recommendations on how to avoid the pitfalls.

The other three aforementioned studies that relate SE  and ML, \cite{breck2017ml, amershi2019software, serban2020adoption}, are the ones that are more closely related to our approach. First, Breck \etal~\cite{breck2017ml}  list 28 practices for testing and monitoring different stages of the ML development process. Amershi \etal~\cite{amershi2019software} conducted a study that reports a broad range of practices and challenges observed  in software teams at Microsoft as they develop AI-based applications. Nevertheless, the set of challenges and practices are broad and often not actionable. They  also focus on a single enterprise (Microsoft).  Finally, Serban \etal~\cite{serban2020adoption} listed  a series of best practices for ML applications.  The presented practices are mined  from academic and gray literature with an engineering perspective, meaning that  the practices are from an engineering point of view and not ML, software engineering for ML (SE4ML).  Serban \etal~\cite{serban2020adoption}  also present a taxonomy of the SE4ML practices, and the taxonomy has  six categories: data, training, coding, deployment, team, and governance. This taxonomy was validated via a  survey in which they asked the respondents (i.e., researchers and practitioners) about the adoption of the  identified  practices on it. The authors also surveyed if adopting a set of practices would lead to a desired effect (\eg agility, software quality, and traceability).  As a result, of their study, they present the list of practices in their article and provide an online tool in which the practices are presented in more detail in the aforementioned taxonomy.

Since our  goal is to help  reduce the gap of not having a clear handbook of ML practices applied to SE, we want to build on the strengths identified in the related work to accomplish that goal. This means that (i) we focus on the approach of ML for SE, trying to understand the perspective of researchers but also practitioners; (ii) give context to the practices and not only the practice itself; in this way, the practices are actionable and meaningful; (iii) related to the last point we want to provide not only case studies in which the practices are used but also help the interested person to identify what task they are trying to achieve/execute in an ML pipeline.

\section{Research Questions }
\label{sec:contr}

The following research questions (RQs) aim to reduce the gap towards having a clear source of ML practices oriented to SE. For that, the RQs are going to be crowdsourced data-driven, which means that the sources of information that will support them are not from a single source of information, but from multiple sources which are not centralized and with different origins. 

\vspace{3pt} 
\begin{quote}\textbf{RQ1.}  \emph{What is the perspective of ML practitioners on best practices, and in which ML stages are they located?}\end{quote}

Answering this research question will help the practitioners' community to understand which stages of the process of developing  ML systems have best practices associated with them, and also could help to avoid pitfalls that are being executed by  omitting, with or without knowing, technical requirements or knowledge of the system.  We will answer this question by studying Stack Exchange posts in which practitioners ask questions about different topics, including ML, as mentioned in previous literature.  
And in order to minimize false positives, we will conduct filtering on the relevant posts based on topic (ML) and quality. After filtering the data, a process of analysis should be carried out in order to extract the possible practices. Subsequently, the practices  should be validated by ML experts in order to filter out the practices that may be outdated or are not considered good practices. 

\begin{quote}\textbf{RQ2.} \emph{What is the perspective  and adoption  of ML practices by researchers and their studies? } \end{quote} 

The identified practices in this research question will give an indication of what practices are being used and reported by the SE research community.  This will help the SE research community to (i) identify possible points to strengthen the research and focus when describing their studies and protocols, (ii) identify possible good practices with SE examples,  which could facilitate the use of good practices and avoid making mistakes. We will answer this question by sampling ML-related papers from SE conferences and identifying ML practices, then they will be categorized in the different ML pipeline stages and SE applications (\ie defect modeling).  Complementary to this, we will conduct a survey and interviews with ML research experts in order to identify their opinion on the use of ML practices and the consequences of omitting them. 

\begin{quote}\textbf{RQ3} \emph{What are the practices identified and adopted by practitioners and researchers? }\end{quote}

Answering this \textit{RQ}  will give both practitioners and researchers a better perspective on the used and identified ML practices. This will help the SE community,  in general,  to be aware of possible practices that are being used and/or omitted. For this \textit{RQ}, we will compile a handbook of practices from the  perspective of  SE researchers and practitioners.  As this \textit{RQ} complements the previous two \textit{RQs}, we will consider the results obtained in \textit{RQ1} and \textit{RQ2} while comparing and complementing the identified practices from both perspectives. In addition, we will enrich the practices with complementary information, such as use cases and previous research, to provide context and examples of their use. Also, we will provide the nature of the perspective (\ie researchers, practitioners, or both). 

\begin{quote}\textbf{RQ4} \emph{To what extend do the identified practices affect previous research?  }\end{quote}

Understanding how the use (or lack of use) of the practices  affects the result of research studies will give the community an idea of the impact and importance that this could cause.  Understanding the impact could generate more awareness of the use and report of the ML practices followed during the study.  For this, we will use a sample of  SE studies that use ML, that can be replicable, which allows us to obtain the same/similar results and then apply or omit ML practices and evaluate how that affects the results that were reported by the studies.  Kindly note that the study will be executed in a way in which, when reporting the results, they will not directly point to specific studies. This study will be a reflective exercise rather than a finger-pointing one, as previously done by the study conducted by Arp \etal~\cite{ArpQuiPen_22} when identifying dos and dont's in ML in computer security.

\section{Current status}  
\label{sec:contr}

This section describes the results achieved so far for each research question in the context of the related work.

Regarding our first research question, in a paper currently under review~\cite{mojicapp}, we used as data source 14 StackExchange websites, including StackOverflow. We decided to use this family of Q\&A because of its popularity in the SE community, which can be seen in multiple studies that have used StackOverflow as a data source to analyze different topics in SE, \eg~\cite{zhang2021study, mondal2023automatic, chatterjee2020finding}. In addition, to its popularity in the SE community, StackOverflow has also been used to study ML-related topics such as expertise and challenges~\cite{Alshangiti_2019}, problems and challenges for ML libraries~\cite{Islam_2019}, and popular deep learning topics~\cite{Han_2020}.


As a result of selecting Q\&A Stack Exchange websites, filtering posts from them to extract the possible ML practices, and analyzing them, we obtained 157 ML best practices and their taxonomy.  The practices were obtained by executing an open-code procedure in which tags for the different practices were identified and assigned together with the identification of the ML pipeline stage(s) associated with each possible practice. For  stage identification, we used a predefined ML pipeline built by Amershi \etal \cite{amershi2019software}. As a result of the open-coding process, a list of 187 practices was identified, but only 157 were considered  best practices by ML experts after a validation process. 

Another outcome of the open-coding process is a four-level taxonomy. The first level of the taxonomy consists of the 10 ML-pipeline stages proposed by Amershi \etal \cite{amershi2019software}. The second level consists of categories  that encompass multiple tasks for each ML pipeline stage, \eg the learning category in the model training stage. The third level of the taxonomy is composed of an action/task that can be performed in each ML stage. The fourth level is the practice itself.

Upon further analysis of the practices, we identified which ML-pipeline stages had the highest number of practices and which ones had the lowest number of practices.  On the one side, the ML pipeline stages with the highest number of identified practices were model training and data cleaning, which could indicate an interest of practitioners in those two stages. The interest could be related to (i) model training being the core of the ML pipeline, as it enables the use of a model; (ii) data cleaning is a stage where  data scientists spend most of their time~\cite{anacondainc_2022}.

On the other side, model deployment, model monitoring, and data labeling are the ones with the lowest number of identified practices. Regarding the low number of practices in model deployment and monitoring, this could be due to  these stages are more  related to the “operations” staff~\cite{LewisGrace2021WAIN} (\ie staff in charge of deploying, operating, and monitoring  ML-enabled systems).  Regarding the low number of practices identified for the data labeling stage, it could be related to the intrinsic nature of this stage.  By this, we mean that this stage is inherently  not mandatory in the ML pipeline, as it is only needed when  ground truth is required, \eg supervised and semi-supervised learning. In addition, sometimes, the data used to train models has already been labeled, which could lead to efforts being focused on other ML phases. 

Another  aspect that we noticed when analyzing the identified practices in the Q\&A websites is that they did not cover some specific topics.  Ethics is an example of a topic that was not discussed/covered by the identified practices.  This could indicate that there is a need to explore other sources of information to find ML-best practices, such as  technical blogs like the one presented by IBM~\cite{ibm_ethics}, which presents  an ethical framework for ML. 

When analyzing the validation of the ML experts, we noticed some  aspects to highlight. Firstly,  the majority of the practices considered good were validated by all the experts, which means that there was unanimous agreement. However, 30 practices were rejected by the experts, as only half or less of the experts considered them valid best practices. After inspecting the practices that the ML experts rejected, we noticed that, in most cases, the opinion was divided. This means that most of the time, half of the experts considered that the practices were not good ones, but the other half considered them good ones. This could mean that those practices, with divided opinions, were not well-known, or without a use case scenario, were not clear. In addition, some practices that were considered contradictory, \ie practices that indicate opposite actions, were agreed as good practices by the experts, which could also be an indicator of the need for more context to present an actionable practice. 

Concerning \textit{RQ2}, we are working on a study in which research-track  articles from top SE conferences are being analyzed in order to understand the reported and used ML practices. In this study, we are also taking as a reference the ML pipeline proposed by Amershi \etal \cite{amershi2019software}, for categorizing the identified practices. In preliminary results, we have found that the least mentioned stages, \ie stages in which few practices were identified, are related to model requirements, model deployment, and model monitoring. The last two were expected as it is not common to describe/execute those two stages in a research study. While the first mentioned stage, model requirements, could be considered the basis of a research study that uses ML, as it could define how the models should be built, and not defining it properly could cause disastrous consequences.

Regarding  \textit{RQ3}, as part of the process of presenting a handbook of practices with both perspectives, practitioners, and researchers, we are currently designing an approach/tool, that will not only be able to be referenced but that will be useful in a practical way. With that, we mean that the practices will be associated and enriched with context, examples, and possible identified limitations.  In addition, this tool should present the aforementioned information in a friendly way, which will allow the users of the tool to find relevant information without going through an entire book, blog, or research article. For that, we are identifying ways  that  practices could be presented in a more interactive way,  like the appendix presented by Serban \etal \cite{serban2020adoption} for the SE practices for ML, the practices presented by Google in their  ``People + AI guided book'' \cite{goole_pair},  the ``Deep Learning Tuning Book''\cite{google_play_book} focused on the process of  model hyperparameter tuning addressed to engineers and researchers. We also take as a reference other white and gray literature aforementioned in the related work that, for each practice, present additional information, such as use cases, \eg~\cite{wujek2016best, ArpQuiPen_22}.

\section{Timeline}
\label{sec:time}

I am currently in my third year of my four-year Ph.D. program. In my third year, I plan to continue working on  my research, focusing mainly on \textit{RQ2} and \textit{RQ3}, while also finishing answering \textit{RQ1}.   In my last year, we will focus on \textit{RQ4} in order to complete it in my fourth year.

\section{Conclusion}
\label{sec:concl}

As a conclusion of this thesis, a synthesis of all four research questions will be provided, including a proposed tool to retrieve the practices.  This will help to reduce the gap of not having a clear handbook of ML practices applied to SE, since the set of validated practices will be oriented to the SE community with practitioners' and researchers' perspectives, which will be complemented with context and SE use cases.

\balance
\bibliographystyle{IEEEtran}
\bibliography{local}

\end{document}